\documentclass[a4paper,11pt]{article}
\pdfoutput=1 % if your are submitting a pdflatex (i.e. if you have
\usepackage{jheppub} % for details on the use of the package, please
\usepackage[T1]{fontenc} % if needed
\usepackage{hyperref}
\usepackage{url}
\usepackage{float}
\usepackage{orcidlink}
\usepackage{slashed} 			 %Feynman slash
\graphicspath{{Figures/}} 	     %Set the default folder for images

\title{\bf  The Radial Mode of Composite Higgs Theories at the LHC}

\author[a,b]{ Gustavo  Burdman\orcidlink{0000-0003-4461-2140},}
\author[a]{Marvin M. Janini\orcidlink{0000-0001-9562-3363},}
\author[a]{Lincoln Pereira\orcidlink{0009-0007-8971-2694}}
\author[a]{and Murilo Trevisan\orcidlink{0009-0000-7661-8606}}

 \affiliation[a]{Department of Mathematical Physics,\\
Institute of Physics, \\
University of Sao Paulo,\\
R. do Matao 1371, Sao Paulo, \\
SP 05508-090, Brazil\\
~~~~~}

\affiliation[b]{Center for Theoretical Physics,\\
  Columbia University,\\
  New York, NY 10027, USA\\
~~~}

\emailAdd{gaburdman@usp.br, marvinjanini@usp.br, linsilpe@usp.br, murilomtrevisan@usp.br}

\abstract{ We examine the potential of the LHC to observe the scalar radial excitation present in extensions of the standard model
  where the Higgs boson is a pseudo Nambu Golstone boson. These 
  include composite Higgs models as well as the twin Higgs model. 
  These states can be  light enough to be seen
  at the LHC, potentially resulting in additional clues about the nature of the
  Higgs sector. We present the current status of LHC bounds as well as
  the future prospects for the   the high luminosity LHC (HL-LHC). 
  We identify the most sensitive channels as those where the radial
  state decays to a pair of Higgs bosons, especially at the high
  luminosity stage. For the minimal composite Higgs models we study, we make
  use of the LHC Run 2 data with ${\cal L}=138~{\rm fb}^{-1}$ to 
  extract the $2\sigma $ mass bounds $m_\sigma\geq (0.93-1.13)~$TeV, where the
  values on the interval depend on the
  parameters of the model. We show that the  reach of the HL-LHC  for  these cases is
  $m_\sigma\geq (1.8-2.2)~TeV$, with ${\cal L}=3000~{\rm
    fb}^{-1}$. For the twin Higgs model radial state, the current
  bounds are set by Higgs coupling measurements, while for the  HL-LHC
  we obtain the reach $m_\sigma\geq 1.2~$TeV, corresponding to the
  lowest symmetry breaking scale allowed by current data. 
}

\begin{document} 
\maketitle
\flushbottom

\section{Introduction}
\label{sec:intro}
Although the dicovery of the Higgs
boson~\cite{Aad_2012,Chatrchyan_2012} completed the standard model
(SM) particle
spectrum, the origin of the Higgs sector 
remains an open question. In particular the origin of the electroweak
scale, the only dimensionful  parameter in the SM Lagrangian, is unknown as well
as highly sensitive to high energy quantum corrections. This
so called hierarchy problem states that the most natural scale for
electroweak symmetry breaking, i.e. for the Higgs boson mass, should
be close to the SM cutoff $\Lambda_{\rm SM}$, the maximum energy where the theory can be
trusted. At the moment, we have experimental knowledge of a {\em  lower
bound} on $\Lambda_{\rm SM}$ from electroweak precision observables as
well as LHC searches~\cite{ParticleDataGroup:2024cfk,Reina:2025suh}
\begin{equation}
\Lambda^{\rm exp.}_{\rm SM} \gtrsim {\rm ~few~TeV}~, 
\end{equation}
where the exact bound depends on the observable. In any case this
already  posses the question known as the little hierarchy problem:
why is it that
\begin{equation}
  m_h \ll \Lambda^{\rm exp.}_{\rm SM}~.
  \label{littleh}
\end{equation}   
A number of extensions of the SM explain (\ref{littleh}) by the fact
that the Higgs boson might be a (pseudo) Nambu-Goldstone boson (pNGB) arising
from the spontaneous  breaking of a global
symmetry~\cite{Kaplan:1983sm,Bellazzini:2014yua}.  It would be this
global symmetry that protects $m_h$ from getting the usual large
corrections that characterize scalar theories.  We divide these pNGB
Higgs  models in two groups: composite Higgs models (CHM)~\cite{Contino:2003ve,Agashe:2004rs} and twin Higgs
models (THM)~\cite{Chacko:2005pe}. In both classes of models the Higgs
is a pNGB, where a spontaneously broken global symmetry explains the
scale separation with $\Lambda^{\rm exp.}_{\rm SM}$.  A relatively
small explicit breaking of the global symmetry, provided by the SM
gauge and Yukawa interactions, is ultimately responsible for the non
vanishing value of  $m_h$.

Both types of models are analogous to the situation in low
energy QCD. In that case, the spontaneously broken global symmetry is
$SU(2)_L\times SU(2)_R\times U(1)_L\times U(1)_R$ broken down to
$SU(2)_V\times U(1)_V$. This results in three NGBs $\pi^\pm$ and
$\pi^0$, which acquire a small mass \footnote{The $U(1)_A$ is anomalous, so there is no NGB
  associated to its spontaneous breaking.}
due to the fact that the global symmetry is explicitly broken by the
quark masses in the QCD Lagrangian.  This guarantees that $m_\pi\ll
\Lambda_{\rm had.}$, with $\Lambda_{\rm had.}\simeq 1~$GeV, the
typical hadronic scale or cutoff of hadron physics. 

Thus, in analogy with low energy QCD, these SM extensions assume that the global
symmetry is spontaneously broken at a scale $f$, which is typically
just above the electroweak scale $v\simeq 246~$GeV. It is then
possible to estimate the cutoff of these theories using a naive
dimensional analysis (NDA)~\cite{Manohar:1983md}  estimate. Then, we obtain that~\cite{Bellazzini:2014yua}
\begin{equation}
\Lambda  \lesssim 4\pi f~.
\end{equation}
In this way, we see that these extensions of the SM would predict the
scale of a new physics threshold {\em beyond} the effective description
of the pNGB, just as in the case of hadronic physics we expect
$\Lambda_{\rm had. }$ to prelude the appearance of the underlying QCD
dynamics of quarks and gluons.

Just as in the case of low energy QCD, where its symmetries dictate
the dynamics of light pions in the presence of a heavier radial
excitation, the CHM and the THM also predict the existence of this 
heavier degree of freedom, which we will generally call the $\sigma$.  The
phenomenology of this state is somewhat different in CHM and THM, so
we will study it separately.

The radial state in theories where the Higgs is a pNGB has been studied in the literature. In the case of
CHMs, Refs.~\cite{Fichet:2016xvs,Fichet:2016xpw} provided a
first comprehensive study. For the case of the mirror twin Higgs models (MTH) Ref.~\cite{Chacko:2017xpd,Ahmed:2017psb}
contains all the elements of the radial state phenomenology. In
Ref.~\cite{Ahmed:2017psb} the study is extended to include the
fraternal twin Higgs model (FTH), in which only the third generation
of fermions is replicated in the twin sector. The phenomenology of the
radial excitation suffer only minor modifications due to the
appearance of new decay modes or the enhancement of certain modes in
the invisible width. These will only result in small modifications of
our results for the MTH, so we will on this simple scenario for the
remaining of the paper.     

In this paper, we update the status of these radial states using data
from the LHC Run 2,  resulting in  bounds on the $\sigma$ mass for the
various scenarios.  
It is of particular interest to
understand the current bounds as well as the future reach of the LHC,
including in the high luminosity era.  In addition, there has been
some reports of interpretation of data~\cite{CMS:2023boe} pointing to the possible
existence of signals for new scalar states~\cite{Crivellin:2021ubm,Khanna:2025cwq,Hmissou:2025riw,Benbrik:2025wkz}. All in all, it is a good
moment to reevaluate the potential of the LHC to see these radial excitations.

Experimental constraints on the THM, come from contributions of the
twin sector to the  invisible
Higgs width~\cite{ATLAS:2023tkt}, whereas for the CHM there are bounds on direct
searches for the associated vector and fermion resonances~\cite{CMS:2024bni,ATLAS:2019lsy,CMS:2019kaf}.
But the most stringent constraints on
 CHM and THM are from the modification of the Higgs boson couplings compared
 with the SM ones~\cite{CMS:2022dwd,ATLAS:2021vrm}, which results in
 the most stringent 
 constraints on the scale of spontaneous symmetry breaking in each case. 
This program of precision measurements of the Higgs couplings is of outmost importance and it
will be continued in the HL-LHC era. However, here we point out that
the possibility of discovery for new scalar states in these theories
is still open, even after these bounds are considered. We will show that  the  Run 2 dataset of ${\cal L} = 138 ~{\rm fb}^{-1}$
integrated luminosity results already in interesting bounds on
$m_\sigma$ (at least in the case of CHM), which shows that there is still
reach for discovery of new states at the LHC. As we will see below, this
will be more so in  the high luminosity runs, accumulating ${\cal L} =
3000~{\rm fb}^{-1}$.  
The identification of the radial mode at the LHC would be a clear signal
for these models. Furthermore, and as we will show below, the
observation of this state, in combination with precise data on Higgs
couplings, could determine which of the two scenarios is realized, as well as providing a complementary test of
these well motivated extensions of the SM beyond the precision
measurements of the Higgs couplings. 

The paper is organized as follows: in Section~\ref{sec:chmradial} we
present the radial state in the context of the minimal composite Higgs
models (MCHMs). Here, we define the details of the models chosen to
study the phenomenology of these theories.
In Section~\ref{sec:radialtwin} we do the same for the radial state of
the TH. We study the phenomenology of the radial state at the LHC  in both
scenarios in Section~\ref{sec:pheno}, and the prospects for the HL-LHC
in Section~\ref{sec:future}. We finally conclude in
Section~\ref{sec:conc}. 

\section{The Radial State in Composite Higgs Models}
\label{sec:chmradial}
In this section we consider the scalar sector of the minimal CHM
(MCHM)~\cite{Agashe:2004rs}. This has been considered previously in
the literature in Refs.~\cite{Fichet:2016xvs,Fichet:2016xpw}. In this
model  a global $SO(5)$ symmetry is spontaneously broken to $SO(4)$.
The minimal scalar content corresponds to embedding the Higgs in a
5-plet of $SO(5)$, resulting in one additional non NGB
scalar~\footnote{As shown in ~\cite{Fichet:2016xvs}, it is possible
 to use a 14-plet of $SO(5)$, resulting in the singlet plus a triplet
 of scalars. If these are considered heavier than the singlet, the
 resulting phenomenology is very similar to the one obtained by
 our present choice.}
The Higgs doublet degrees of freedom emerge as the 4 NGBs associated
with this breaking. 
The Lagrangian describing the scalar sector can be written as~\cite{Panico:2015jxa}
\begin{equation}
{\cal L} = \frac{1}{2}(D_\mu\Phi)^\dagger D^\mu\Phi
-\frac{g^2_*}{8}\left(\Phi^T\Phi - f^2\right)^2~~,
\label{philag}
\end{equation}
where $D_\mu$ refers to the $SU(2)_L\times U(1)_Y$ electroweak
covariant derivative
\begin{equation}
D_\mu\Phi = (\partial_\mu -ig W_\mu^\alpha T_L^\alpha  -ig' B_\mu
T_R^3)\Phi ,
\label{covdersm}
\end{equation}
and $\alpha=1,2,3$. In the expression above
\begin{equation}
T^\alpha_L =\left(\begin{array}{cc} t^\alpha_L &0\\0&
                                                      0\end{array}\right)
                                                  \qquad
T^3_R =     \left(\begin{array}{cc} t^3_R &0\\0&
                                                      0\end{array}\right)~,                                              
                                                \end{equation}
represent the                                                 
three generators
of $SU(2)_L$ and the one generator of $SU(2)_R$ that are gauged,
written as embedded in $SO(4)$ in terms of their usual $2\times 2$ form
$t_L^\alpha$ and $t_R^3$, and 
indicating the partial gauging of $SO(4)\sim SU(2)_L\times
SU(2)_R$. The 5-plet $\Phi$ can be written as 
\begin{equation}
\Phi(x) = e^{i\sqrt{2}\pi^{\hat a}(x) t^{\hat
    a}/f}\left(\begin{array}{c}\mathbf{0}\\~\\
                 \sigma(x)+f\end{array}\right)
             \label{phidef}
\end{equation}
where the $t^{\hat a}$ correspond to the $SO(5)/SO(4)$ coset (broken)
generators and $\pi^{\hat a}(x)$ are the NGB degrees of freedom. The
radial excitation, $\sigma(x)$ will be the main object of this
paper. It is in principle always present when a continous symmetry is
spontaneously broken, and the MCHM is no exception. However, it can be
argued that the radial excitation, since is considerably heavier than the
pNGBs, may be ignored (integrated out) as long as we consider
processes with momenta well below its mass
\begin{equation} 
  m_\sigma = g_* f~.
  \label{sigmamass}
\end{equation}
To a large extent, this is the case in low energy QCD, where $m_\pi
\ll m_\sigma\simeq 500~$MeV~\cite{ParticleDataGroup:2020ssz}. Furthermore, the radial excitation is
very strongly coupled resulting in $\Gamma_\sigma^{\rm QCD}\simeq
m_\sigma$, making its observation more difficult. Experimentally, the
presence of the $\sigma$ in low  energy pion scattering experiments is
determined through the $\pi-\pi$ phase shift, which goes through
$\pi/2$ at $s=m_\sigma$.   In the present case, we will study the
region of parameter space available to the LHC. This will imply a
radial excitation with a raltively smaller width. This need not apply
to the vector resonances, which are governed by a different coupling
$g_\rho$, and a different scale $f_\rho$~\cite{Fichet:2016xvs}. We will be exploring the
phenomenology of the radial excitation in the limit $f_\rho\gg f$,
which in turn implies that $f$ and $g_*$ are the only paramenters of
interest. We will comment on the effects of this approximmation on our
results when relevant. 

In order to read off the couplings of $\sigma$ to the Higgs boson, it
is advantageous to rewrite $\Phi$ as
\begin{equation}
\Phi(x) = (f+\sigma(x))\, \left(\begin{array}{c}
                                  \frac{\mathbf{\Pi}}{\Pi}\sin(\frac{\Pi}{f})\\~\\
                                  \cos(\frac{\Pi}{f})\end{array}\right)
                              ~,
\label{phifield}                              
\end{equation}   
where $\mathbf{\Pi}$ is a $\bf{4}$ of $SO(4)$, and we defined
$\Pi\equiv \sqrt{\sum_i\Pi_i^2}$. This, in turn, can be
decomposed in terms of the Higgs doublet by writing the NGB 4-plet
as $\mathbf{4} = (2,2)$ in terms of the Higgs doublet and its
conjugate.
Defining the $SU(2)_L$ Higgs doublet as 
\begin{equation}
H = \left(\begin{array}{c} \varphi_u \\ \varphi_d\end{array}\right), 
\end{equation}
in terms of the complex scalars $\varphi_u$ and $\varphi_d$, we arrive at 
\begin{equation}
\mathbf{\Pi} = \left(\begin{array}{c} \Pi_1\\ \Pi_2\\ \Pi_3\\
                       \Pi_4\end{array}\right) = \frac{1}{\sqrt{2}}
                     \left(\begin{array}{c}
                             -i(\varphi_u-\varphi_u^\dagger) \\
                             \varphi_u+\varphi_u^\dagger\\
                             i(\varphi_d-\varphi_d^\dagger)\\ \varphi_d+\varphi_d^\dagger\end{array}\right)
\end{equation}
Next, we go to the unitary gauge, where $\varphi_u=0$ and
\begin{equation}
  \varphi_d =\frac{v+h(x)}{\sqrt{2}}~,
\end{equation}
resulting in $\mathbf{\Pi}^T = (0 ~0 ~0~(v+h(x))$ .
We obtain the couplings of $\sigma$ to the Higgs boson, and to the SM longitudinal
gauge bosons from
\begin{eqnarray}
{\cal L} &\supset& 
\left(1+\frac{\sigma}{f}\right)^2\,\left\{\frac{1}{2}\partial_\mu h\partial^\mu
  h+\frac{g^2 f^2}{4}
  \,\sin^2\left(\frac{v+h}{f}\right)\,\left(W^+_\mu W^{-\mu}
                   + \frac{1}{2 \cos\theta_W^2} Z^\mu
                   Z_\mu\right)\right\} \nonumber\\
  && ~\nonumber\\
  &\supset&  \frac{1}{f} \sigma \partial_\mu h\partial^\mu h+
            \frac{2M_W^2}{f} \,\sigma\,W^+_\mu W^{-\mu} + \frac{M_Z^2}{f} \,\sigma\,Z^\mu
            Z_\mu + \cdots ~.
            \label{sigmacouplings1}
\end{eqnarray}
 It is then straightforward to
compute the partial decay widths of $\sigma$ to the Higgs, $Z$ and
$W^\pm$ bosons. This results in
\begin{equation}
\Gamma(\sigma\to h h) \simeq
\Gamma(\sigma\to ZZ) \simeq \frac{1}{2}\,\Gamma(\sigma \to W^+
W^-)~\simeq \frac{m_\sigma^3}{32\pi f^2},
\label{sigmawidths}
\end{equation}
up to correction that go like $m^2_h/m_\sigma^2$, $m_Z^2/m_\sigma^2$,
$m_W^2/m_\sigma^2$ or smaller. The decay mods above will dominate the
phenomenology of the radial excitation $\sigma$ at the LHC. 

\subsection{Fermion Couplings}
\label{subsec:fermioncouplings}
The couplings of the radial excitation to fermions in the MCHM are more model
dependent since they introduce new parameters, and the fermion
representation in which the SM fermions are embedded must be chosen.
This choice is informed by the so called partial compositeness
framework~\cite{Kaplan:1991dc,Contino:2004vy,Contino:2006qr,Csaki:2008zd}. Elementary fermions with the SM quantum numbers are
external to the composite sector. They couple to it via linear couplings to operators with the
appropriate quantum numbers~\cite{Panico:2015jxa}. In turn, these
operators excite composite sector resonances with these same quantum
numbers. For instance,
\begin{equation}
\langle 0|{\cal O}^L | Q\rangle \not=0, \qquad \qquad \langle 0|{\cal O}^R | T\rangle \not=0~,
\end{equation}
where we consider here operators that may couple linearly to an
elementary quark doublet $q_L$, or to an elementary singlet such as
$t_R$. Thus, composite Higgs models contain resonances with the SM
quantum numbers. In fact, since the Higgs is composite, it only
couples to the elementary sector trough the resonances which mix with
it. Our interest here is to obtain the couplings of the resonances to
the radial excitation $\sigma$. For this purpose, we must write the
lowest dimensional operators coupling $\Phi$ with these states and
respecting the $SO(5$ symmetry. However, in order to do this, we must
first choose the representation of $SO(5)$ for the resonances. This
introduces an additional model dependence: choosing the fermion
representation such that we can correctly embed the SM fermions.

We will consider as examples cases where the scalar $\Phi$ is always
introduced in the fundamental representation of $SO(5)$, i.e. as a
$\mathbf 5$ od $SO(5)$. 
On the other hand, we need to choose fermion
representations in order to embed the elementary fermions, This choice will be important since 
the fermion content includes heavy resonances that affect the loop
induced couplings of  the radial mode $\sigma$, which will partially determine  
some of its decays widths and most crucially its production through
gluon fusion.
In order to obtain the correct hypercharges for the SM fermions, which
will be embedded in those $SO(5)$ representations, we need to extend
the symmetry to include a global $U(1)_X$, i.e the symmetry is now
$SO(5)\times U(1)_X$.
There are many fermion representations that would allow us
to build $SO(5)\times U(1)_X$  invariant Yukawa couplings, even
restricting ourselves to the case where $\Phi$ is a $\mathbf 5$ of $SO(5)$. Since our aim is to be
illustrative, as opposed to exhaustive, we will only consider
dimension four couplings operators.   Furthermore, we restrict our
choice imposing that the elementary $q_L$ be embedded in a
$\mathbf{5} = (\mathbf{2},\mathbf{2})+\mathbf{1}$ so that the
bidoublet under $SO(4)\sim SU(2)_L\times SU(2)_R$ respects a custodial
symmetry necessary to avoid large deviations in $Z\to b_L\bar b_L$. We
these considerations, we settle for the following  fermion
representations for the $Q_i,U_i,D_i$ resonances, where $i$ is a
generation index:
\begin{itemize}
\item  ${\rm {\bf MCHM}}_{\mathbf{5,1,10}}$: the vector-like excitations of $q^i_L$ are
  in a $\mathbf{5}_{2/3}$ representation of $SO(5)\times U(1)_X$,
  whereas the excitations of $u_R^i$ and $d_R^i$ are, respectively, in
  $\mathbf{1}_{2.3}$ and $\mathbf{10}_{2/3}$ representations. The
  Yukawa couplings can be written as~\cite{Fichet:2016xvs}
  \begin{equation}
{\cal L}_Y = - \xi_U^{ij} \bar Q_{i} \Phi P_R U_j - \xi_D^{ij} \bar Q_i
P_R D_j \Phi -\xi_U^{'ij} \bar Q_i \Phi P_L U - \xi_D^{'ij} \bar Q_i
P_L D_j\Phi +  {\rm h.c.}, 
\label{mchm5110}
\end{equation}
where $P_{L,R}$ are chirality projectors. We see that there are two
independent proto Yukawa operators that can be written, leading to the
$\xi_{U,D}$ and $\xi'_{U,D}$ matrices. 

\item ${\rm{\bf MCHM}}_{\mathbf{5,14,10}}$: In this case, the up sector is
  represented by a $\mathbf{14}_{2/3}$ of $SO(5)\times U(1)_X$, with
  $Q_i$ and $D_i$ still in the $\mathbf{5}_{2/3}$ and
  $\mathbf{10}_{2/3}$ representations. The Yukawa operators then take
a similar form 
  \begin{equation}
{\cal L}_Y = - \xi_U^{ij} \bar Q_{i}  P_R U_j\Phi  - \xi_D^{ij} \bar Q_iP_R
D_j \Phi -\xi_U^{'ij} \bar Q_i P_L U\Phi - \xi_D^{'ij} \bar Q_i
P_L D_j\Phi + {\rm h.c.} .
\label{mchm51410}
\end{equation}
 \end{itemize} 
The proto Yukawa matrices $\xi_U$,  $\xi_D$, $\xi'_U$ and $\xi'_D$  above are assumed to be
of order one, consistent with the so called anarchical ansatz. This
involves the assumption that the flavor hierarchy reflects the
composite character of the fermions, as defined by their mixing with
the resonances of the composite sector: those with larger Yukawas have
a larger mixing, whereas those that are more elementary-like and
therefore lighter have small mixings with the strong sector.

Alternatively, it is possible to consider a hierarchical approach,
where the proto Yukawas above are chosen so as to produce directly the SM
flavor spectrum, whereas all the mixings are chosen to be of the same
order. This case is simpler that the one shown above, and sometimes
goes by the name ${\rm \bf MCHM_{\mathbf{5,1}}}$, given that we need
only specify the fermion resonances associated with the top quark. 
Here, we will only consider the anarchical case, not just due to its
appeal but, as we will see later, it will result in a wider range of
masses for the radial state that are amenable to searches at the LHC.

Finally, we
need to compute the top quark coupling to the radial excitation
$\sigma$ which will also enter crucially in the production
mechanism. This coupling arises only after electroweak symmetry
breaking. In order to extract it we consider the mixing Lagrangian,
in which we can include the $\sigma$ coupling, appearing always in the
$f+\sigma(x)$ combination. From the mixing terms we can diagonalize
the mass matrix. Then, the physical top mass and couplings to $\sigma$
can be determined ~\cite{Panico:2015jxa}
\begin{equation}
{\cal L}_{\rm top} \supset (f+\sigma(x)) \frac{v}{f} \,{\cal C} \bar
t_Lt_R + {\rm h.c.}~, 
\end{equation}
where ${\cal C}$ depends on the $SO(4)$ masses and the mixing
parameters. As a result, we see that the top couples to the radial
mode $\sigma$ with strength ~\cite{Fichet:2016xvs}
\begin{equation}
 \frac{m_t}{f}~.
  \label{top2sigma}
\end{equation}

We now have all the elements needed to compute the production and
decay of the radial state $\sigma$ at the LHC. One last step is to
compute the loop induced $\sigma$ couplings to gauge bosons, including
photons, $W^\pm$'s and $Z$'s, as well as gluons. In particular, the
latter will be central for computing the production cross section, so
we focus on it here.
We start with the effective gluon-gluon-$\sigma$ coupling given by
\begin{equation}
{\cal L}_{gg\sigma} = -\frac{c_{gg}}{f}\,\sigma G^a_{\mu\nu}G^{a\mu\nu}
~.
\label{leffggsigma}
\end{equation}  
The coefficient $c_{gg}$ receives contributions from both the top quark
loop as well as contributions from the strong sector that are integrated
out. We will approximate these last ones by including strong sector
fermionic resonances, which will introduce model dependence associated
with the choice of representation for them.

The top contribution is given by
\begin{equation}
c^t_{gg} =
\frac{\alpha_s}{16\pi}\,A_{1/2}\left(\frac{m_\sigma^2}{4m_t^2}\right)~,
 \label{toploop} 
\end{equation}  
where $\alpha_s$ is the strong interaction coupling and $A_{1/2}(x)$
is the standard loop function given, for instance,  in page 439 of Ref.~\cite{Donoghue:2022wrw}. 

Next, we compute the contributions of fermionic resonances. They are
given by
\begin{equation}
c^\psi_{gg} = \frac{\alpha_s}{8\pi}\frac{f^2}{M^2_\psi}
\,A_{1/2}\left(\frac{m_\sigma^2}{4 M^2_\psi}\right) \,N_\psi~,
\label{resloop}
\end{equation}
where we assumed that
$M_Q=M_U=M_D\equiv M_\psi$ is a common resonance mass, and $N_\psi$ encodes the specific information of the resonance
sector in the chosen representation.This includes both the
multiplicity as well as the traces over the proto Yukawa matrices.
We can write it as
\begin{equation}
N_\psi = N^U\, {\rm Tr} \xi^{'T}_U \xi_U +N^D\,{\rm Tr} \xi^{'T}_D
  \xi_D~,
\label{nupanddown}
\end{equation}  
with the $N^U$ and $N^D$ the multiplicities corresponding to the given
representation of the fermion resonances. 
For our two examples we have~\cite{Fichet:2016xvs}:
\begin{eqnarray}
  {\rm \bf MCHM}_{\mathbf{5,1,10}}:&&  \quad N^U=1\qquad\qquad N^D =2 \label{mchm5110_2}\\
  ~~&& \nonumber\\
  {\rm \bf MCHM}_{\mathbf{5,14,10}}:&&\quad N^U=\frac{14}{5}\qquad\qquad N^D=2 \label{mchm51410_2}
\end{eqnarray}
We note that in (\ref{nupanddown}) the products $\xi_U^{'T}\xi_U$ and
$\xi_D^{'T}\xi_D$ appear, signaling that both chiral projections of
$U$ and $D$ must be present in equations (\ref{mchm5110}) and
(\ref{mchm51410}) in order for the resonances to contribute to the
loop diagram entering in the effective coupling $c^\psi_{gg}$. This is
due to the fact that if the $\sigma$ vertex in the triangle diagram comes from
one of the operators with a given chiral projection, the insertion of
$f$ in the propagator between the two gluon vertices  must come from
the other projection for the fermion chirality flow to work. 

In order to further simplify the calculations, we will assume that the
proto Yukawa matrices satisfy $\xi_U=\xi'_U=\xi_D=\xi'_D=\xi$. This resonance
Yukawa coupling will then be one of the parameters we will use in the
phenomenology of $\sigma$ at the LHC.

In sum, the choice of fermion representations to study is greatly
informed by the need to respect $SU(2)$ custodial symmetry, as well as
to reflect the anarchical Yukawa approach consistent with the idea of
partial fermion compositeness. Then, we are left with the   $ {\rm \bf
  MCHM}_{\mathbf{5,1,10}}$ and $ {\rm \bf MCHM}_{\mathbf{5,14,10}}$
fermion representations described above. The one representation that
is still viable in terms of custodial symmetry, but corresponds to a
hierarchical set of fermion couplings (contrary to the partial
compositeness principle) is the ${\rm \bf MCHM}_{\mathbf{5,1}}$, where
  only the top sector is relevant. As we will see immediately below,
  when taking into account running effects to define the paramenter
  space for the mass $m_\sigma$, we will see that in this case this
  mass results to be quite heavy, and therefore of less interest for
  the LHC phenomenology considered here, i.e. that of the direct
  search of the radial mode. Our final simplification is to consider
  the proto-Yukawa couplings in (\ref{mchm5110}) and
  (\ref{mchm51410}), which in accordance to the partial compositeness
  requirement must be all of order one, to be exactly the same and
  equal to $\xi$. This simplification is not likely to have any
  significant impact on the  phenomenology of the radial state, even
  when considering the running effects (see below) given that partial
  compositeness already imposes that their values should be of the
  same order. The flavor structures in these models are built from the
  variety of compositeness, i.e. the very different mixings of composite
  and elementary states, not from the proto-Yukawa couplings.   

\subsection{Running effects and the allowed values of $m_\sigma$ in
  the MCHM}
The coupling $\xi$ defined above will be subjected to renormalization
running. This, in turn, will affect the running of the coupling $g_*$ defined
in (\ref{philag}). Then, imposing constraints on these values rooted
on the stability of the potential, as well as on unitarity, will
result in constraints on $m_\sigma$ as defined in
(\ref{sigmamass}). Following Ref.~\cite{Fichet:2016xvs}, we define
\begin{equation}
\xi_{\rm eff.}^2  =  4 N_c \left( N^U  \left[{\rm Tr}\xi_U\xi^T_U +
    {\rm Tr}\xi'_U\xi^{'T}_U\right]  + N^D \left[{\rm Tr}\xi_D\xi^T_D +
    {\rm Tr}\xi'_D\xi^{'T}_D\right] \right),
\end{equation}
corresponding to the wave function renormalization of the
$\sigma$, which dominates the running of the Yukawa couplings, due to
the enhanced effects of the multiplicity of resonance fermions. The
renormalization group equation is approximately given by
\begin{equation}
\mu\frac{d\xi^2_{\rm eff.}}{d\mu} \simeq \frac{1}{16\pi^2} \xi^2_{\rm
  eff.}.
\label{rge1}
\end{equation}
On the other hand, the $\Phi$ self coupling defined in
(\ref{philag}), obeys the RGE
\begin{equation}
\mu\frac{d g_*^2}{d\mu} = \frac{1}{16\pi^2} \left( \frac{13}{2} g_*^4 + 2g_*^2
  \xi^2_{\rm eff.} - 2\epsilon \xi^2_{\rm efff.}\right)~,
\label{rge2}
\end{equation}
where we defined the coefficient $\epsilon$ to cancel the excess
multiplicity in $\xi^4_{\rm eff.}$ with respect to the actual
resonance loop contributing to the quartic coupling, and it depends on
the fermion representation chosen. In order to solve
these equations, we first impose on (\ref{rge1}) that, at the cutoff
$\Lambda$, $\xi_{\rm eff.}(\Lambda) = 4\pi$. We then solve
(\ref{rge2}) by imposing two types of boundary conditions at
$\mu=\Lambda$.  In the first case, we want to know the {\em maximum} value of
$g_*(m_\sigma)$  that will not result in a Landau pole before the
cutoff $\Lambda$, i.e. the evolution that would result in $g_*(\Lambda)=4\pi$. 
In the second case, we want to know the {\em minimum}
value of $g_*(m_\sigma)$ that would still not result in
$g^2_*(\Lambda)<0$, turning the potential unstable.
We  consider  the cutoff to be either $\Lambda=3m_\sigma$ or
$\Lambda=4m_\sigma$. These are suitably conservative choices, that will
stay considerably below the overall cutoff of the composite theory,
$4\pi f$. The dependence on the particular choice for $\Lambda$ is
very mild and only noticeable for maximum values of $g_*(m_\sigma)$. 
The results for the two chosen fermion representations are shown
in Table~\ref{table1}.

We finally briefly comment on the minimal hierarchical ansatz
${\rm\bf MCHM}_{\mathbf{5,1}}$, where only the top quark Yukawa
  contribution  appears. In this case, the effects of running from the
  cutoff down to $m_\sigma$ are smaller, which results in a minimum
  value for the coupling~\cite{Fichet:2016xvs} $g^{\rm min.}_*(m_\sigma)\simeq 2.1$. Then,
  we see than in addition to this scenario being less theoretically
  appealing, it also results in a smaller range of allowed masses for
  $\sigma$ that
  already start at quite high values, making this choice also
  unappealing phenomenologically. 

With the choices shown in Table~\ref{table1}, and using (\ref{sigmamass}) we
obtain the allowed interval for the values  of $m_\sigma$ for each
case, which we will consider for the phenomenology. 
\begin{table}
  \centering
  \vskip0.1in
  \begin{tabular}{|c|c|c|c|}
    \hline\hline
Representation & $\epsilon$ & $g^{\rm min.}_*(m_\sigma)$ & $g^{\rm  max.}_*(m_\sigma)$
    \\
    \hline
    ~&~&~&\\
     ${\rm \bf MCHM}_{\mathbf{5,1,10}}$ & $\frac{1}{162} $ & $0.69
                                                             (0.68)$ &
                                                                       $2.99
                                                                       (2.54)$
    \\
   ~ &~&~& \\
    \hline
    ~&~&~& \\
     ${\rm \bf MCHM}_{\mathbf{5,14,10}}$ & $\frac{11}{3456}$ & $0.5
                                                               (0.49)$
                                                         & $2.97
                                                           (2.51)$ \\
    ~&~&~&\\
    \hline
  \end{tabular}
   \caption{Minimum and maximum values of the coupling $g_*$, for each
    choice of fermion representations. The numbers in the third and
    fourth columns correspond to $\Lambda=3m_\sigma$, while the ones in
    parentheses are for $\Lambda=4m_\sigma$. For more details, see
    text.}
  \label{table1}
\end{table}

\section{The Radial State in Twin Higgs Models}
\label{sec:radialtwin}
In the Mirror Twin Higgs model~\cite{Chacko:2005pe}, the Higgs is also
a pNGB of a spontaneously broken global symmetry,  $SU(4)\to SU(3)$.
But unlike in CHMs, the MTH involves a mirror copy  of the SM,
supplemented by a $Z_2$ exchange symmetry that guarantees that
quadratic divergencies cancel. However, the $Z_2$ symmetry cannot be
exact or the scale of electroweak symmetry breaking should be the same
as the scale of the global symmetry breaking $f$. This is excluded
since it would result in a Higgs boson invisible branching fraction
of exactly $50\%$, as well as in suppressed couplings of the Higgs to
all SM states beyond what is allowed by
experiment~\cite{CMS:2022dwd,ATLAS:2021vrm}.
As we will see below, the experimental extraction of the Higgs couplings results in bounds
on the scale $f$. In order to preserve the cancellation of
quadratic divergences in the Higgs boson two point function, the $Z_2$
symmetry is only  softly broken, preserving its validity for
dimensionless couplings such as gauge and Yukawa couplings.

The field $\Phi$ now transforms in the fundamental of $SU(4)$, and it
can be written as
\begin{equation}
  \Phi(x) = \left(\begin{array}{c} H_A \\H_B\end{array}\right)~,
\end{equation}
where $H_A$ and $H_B$ transform as doublets of $SU(2)_A$ and
$SU(2)_B$, respectively. But in order to expose the NGB and the radial
degrees of freedom separately, we can write $\Phi$ analogously to
(\ref{phidef}) in CHMs\footnote{Here we follow the MTH literature in
  choosing the normalization for $f$, which differs from the one in
  (\ref{phidef}) for the MCHM.}
\begin{equation}
\Phi(x) = e^{i\Pi(x)/f}\left(\begin{array}{c}\mathbf{0}\\~\\ \frac{\tilde\sigma(x)}{\sqrt{2}}+f\end{array}\right)   
\end{equation}
where $\Pi(x)=\pi^{\hat{a}}t^{\hat{a}}$, and the $\pi^{\hat{a}}(x)$
are the NGBs, corresponding to the broken generators $t^{\hat{a}}$,
and $\tilde\sigma(x)$ is the radial mode. If we only write the $A$ sector (i.e. the SM sector) NGBs, since the $B$
sector ones will be ``eaten''  by the $W^\pm_B$ and $Z_B$ , we have  
\begin{equation}
  \Pi(x) = \left(\begin{array}{cccc}
                   0& 0& 0& i\tilde h_1 \\
                   0& 0& 0& i\tilde h_2 \\
                   0& 0& 0& 0\\
                   -i\tilde h^*_1 & -i\tilde h^*_2 & 0 & 0 
                   \end{array}\right)
 \end{equation} 
where $\mathbf{\tilde h} = (\tilde h_1 ,  \tilde h_2)^T$ is the SM Higgs
doublet. In this way, the $SU(4)$ field can be written in terms of all
the relevant degrees of freedom as~\cite{Burdman:2014zta} 
\begin{equation}
\Phi(x) =\left(f+\frac{\tilde\sigma(x)}{\sqrt{2}}\right)\, \left(\begin{array}{c}
                                                         \frac{\mathbf{\tilde
                                                         h}}{\sqrt{\tilde
                                                         h^\dagger
                                                         \tilde h}}
                                                         \sin\left(\frac{\sqrt{\tilde
                                                         h^\dagger\tilde
                                                         h}}{f}\right)
                                                        \\~\\
                                                         0\\~\\
                                  \cos\left(\frac{\sqrt{\tilde
                                                         h^\dagger\tilde
                                                         h}}{f}\right)\end{array}\right)
                              ~.
\label{phifield_twin}                              
\end{equation}
If we further impose the unitary gauge in the SM Higgs sector by
choosing $h_1=0$ and $h_2=(v+\tilde h(x))/\sqrt{2}$, we can express
the doublets $H_A$ and $H_B$ as
\begin{equation}
  H_A = \left(\begin{array}{c} 0 \\
                \left(f+\frac{\tilde\sigma}{\sqrt{2}}\right)
                \sin\left(\frac{v+\tilde h}{\sqrt{2}f}\right)\end{array}\right),
\label{ha}
\end{equation}
and
\begin{equation}
  H_B = \left(\begin{array}{c} 0 \\
                \left(f+\frac{\tilde\sigma}{\sqrt{2}}\right)
                \cos\left(\frac{v+\tilde h}{\sqrt{2}f}\right)\end{array}\right).
\label{hb}
\end{equation}
The fields  $\tilde\sigma(x)$ and $\tilde h(x)$ are not the
  physical degrees of freedom since they will mix in the presence of
  soft explicit  $Z_2$ breaking. Following ~\cite{Chacko:2017xpd}, we
 write the potential as
 \begin{eqnarray}
 V &=& -\mu^2 \left(H_A^\dagger H_A + H_B^\dagger H_B\right) +
 \lambda\left(H_A^\dagger H_A + H_B^\dagger H_B\right)^2
       \nonumber\\
   & & +m^2\left(H_A^\dagger H_A - H_B^\dagger H_B\right)
       +\delta\left\{ \left(H_A^\dagger H_A\right)^2 +
       \left(H_B^\dagger H_B\right)^2\right\}~.
       \label{potential}
 \end{eqnarray}
 The first two terms above respect both the global $SU(4)$ as well as
 the $Z_2$ exchange symmetry $A\leftrightarrow B$. The third term
 breaks both symmetries, whereas the last one, breaks $SU(4)$ but
 respects the $Z_2$. The $m^2$ and $\delta$ terms lead to mixing of
 the radial mode $\tilde\sigma$ and the Higgs $\tilde h$. As is
customary in TH models, we define the electroweak scale by
 \begin{equation}
   v_{\rm EW} \equiv \sqrt{2} f \sin\left(\frac{v}{\sqrt{2}
       f}\right) = \sqrt{2} f \sin\theta~.
 \label{vewdef}  
\end{equation}
Using the equations of motion for $\tilde\sigma$ and $\tilde h$, we
can eliminate $m^2$ and $\mu^2$ through the
relations~\cite{Chacko:2017xpd}
  \begin{equation}
   m^2 = \delta\,f^2\,\cos 2\theta~,\qquad\qquad \mu^2 =
   f^2(2\lambda+\delta).  
 \end{equation}
 Then, the potential contains the terms
 \begin{eqnarray}
  V &\supset& \left(2\lambda f^2 + \delta
   f^2(1+\cos^22\theta)\right)\,\tilde \sigma^2 + \delta f^2
              \sin^22\theta\,\tilde h^2 \nonumber\\
    &&  -\delta f^2 \sin 4\theta\,\tilde\sigma \tilde h ~,
       \label {vquadratic}
 \end{eqnarray}
 leading to mixing between $\tilde\sigma$ and $\tilde h$. We define
 the mass eigenstates $\sigma$ and $h$ by the transformation
 \begin{equation}
\left(\begin{array}{c} h\\ \sigma\end{array}\right) =
\left(\begin{array}{cc} \cos\alpha & \sin\alpha \\-\sin\alpha &
                                                                 \cos\alpha\end{array}\right)
                                                             \,
                                                             \left(\begin{array}{c}
                                                                      \tilde
                                                                      h\\ \tilde\sigma\end{array}\right)~.
\label{rotation}                                                             
 \end{equation}  
 The mass eigenstates have masses given by
 \begin{eqnarray}
 m^2_h &=& 2f^2\left\{\delta+ \lambda -
           \sqrt{\lambda^2+\delta(\delta+2\lambda)\cos^2 2\theta}\right\}
   \label{higgsmass}\\
 m_\sigma^2 &=& 2f^2\left\{\delta+ \lambda +
                \sqrt{\lambda^2+\delta(\delta+2\lambda)\cos^2 2\theta}\right\}
 \label{psigmamass}               
 \end{eqnarray}
From these, it is possible to obtain a constraint on $m_\sigma$ that
guarantees that the quartic coupling $\lambda$ stays real. This is
given by~\cite{Chacko:2017xpd}
\begin{equation}
  M_{\sigma}\geq m_h\,\cot\theta~,
\label{msigmabound}
\end{equation}
where it was used that $f/v>1$.

The couplings of the physical radial mode $\sigma$ to SM states can
now be derived.  For instance, the couplings to gauge bosons are
derived from the kinetic terms
\begin{equation}
{\cal L}_{\rm kin.} = |\left(D^A_\mu H_A\right)|^2 +   |\left(D^B_\mu
  H_B\right)|^2~,
\label{lkinetic}
\end{equation}
where $D_\mu^A$ and $D_\mu^B$ refer to the covariant derivatives for
the $SU(2)_A\times U(1)_A$ and the $SU(2)_B\times U(1)_B$
respectively. For instance,
\begin{equation}
D_\mu^A H_A = \left(\partial_\mu -ig W^c_{\mu A} t^c + -i\frac{g'}{2}
  B_{\mu A}  \right) H_A~,
\end{equation}
and similarly for $A\to B$. Expanding the fields in (\ref{ha}) and
(\ref{hb}), and replacing $\tilde\sigma$ and $\tilde h$ with the
physical states as defined by (\ref{rotation}), we obtain the
interactions of $h$ and $\sigma$ with the visible $A$ and the
invisible $B$ gauge bosons. Of particular interest for us here are the
$\sigma$ interactions with $A$ sector gauge bosons.
For instance the coupling with the visible $W^\pm$'s is given by
\begin{eqnarray}
{\cal L}_{\rm kin.} \supset&& \frac{g^2}{2} \sqrt{2} f \sin\theta
\left(\cos\alpha\sin\theta - \sin\alpha\cos\theta\right)\,\sigma
                              W^+_{A\mu} W_A^{-\mu}+\dots\nonumber\\
  &=&2\frac{M^2_{W_A}}{v_{\rm EW}}\,\sin(\theta-\alpha)\,\sigma\,
      W^+_{A\mu} W_A^{-\mu} +\dots,
      \label{sigma2ws}
\end{eqnarray}
where we used $M_{W_A} = gv_{\rm EW}/2$. 
We see that the interactions of the physical radial mode $\sigma$ with
$A$ sector gauge bosons are suppressed by a factor of
$\sin(\theta-\alpha)$, where the sign of the mixing angle $\alpha$
entering here is a result of the choice made in (\ref{rotation}).
The complete couplings of the radial mode to gauge bosons are
\begin{eqnarray}
 &&2\frac{\sigma}{v_{\rm EW}}\,\sin(\theta-\alpha)\,
      \left\{M^2_{W_A} W^+_{A\mu} W_A^{-\mu} +\frac{1}{2} M^2_{Z_A}
  Z_A^\mu Z_{A\mu}\right\}\nonumber\\
  &+&2\frac{\sigma}{v_{B}}\,\cos(\theta-\alpha)\,
      \left\{M^2_{W_B} W^+_{B\mu} W_B^{-\mu} +\frac{1}{2} M^2_{Z_B}
  Z_B^\mu Z_{B\mu}\right\}
  \label{completescouplings}
\end{eqnarray}
whereas the physical Higgs boson couplings are given by
\begin{eqnarray}
 &&2\frac{h}{v_{\rm EW}}\,\cos(\theta-\alpha)\,
      \left\{M^2_{W_A} W^+_{A\mu} W_A^{-\mu} +\frac{1}{2} M^2_{Z_A}
  Z_A^\mu Z_{A\mu}\right\}\nonumber\\
 &+&2\frac{h}{v_{B}}\,\sin(\theta-\alpha)\,
      \left\{M^2_{W_B} W^+_{B\mu} W_B^{-\mu} +\frac{1}{2} M^2_{Z_B}
  Z_B^\mu Z_{B\mu}\right\}~,
  \label{completehcouplings}
\end{eqnarray}
where we defined $v_B\equiv \sqrt{2} f \cos\theta$.
Similarly, we can derive the couplings of $\sigma$ to fermions by
expanding the Yukawa terms. For instance, for the quarks   we have
\begin{eqnarray}
{\cal L}_{\rm Yuk.} = - \bar Q_A \tilde H_A Y_u u_A  - \bar Q_A H_A
  Y_d d_A  + (A\to B) ~,
  \label{yukawas}
\end{eqnarray}  
where $Y_u$ and $Y_d$ are the up and down Yukawa matrices, identical
in both sectors due to the $Z_2$ symmetry. Once the Yukawa matrices
are diagonalized and we obtain fermion masses as
\begin{equation}
m_{q_A} = \frac{Y_{q_A}}{\sqrt{2}} v_{\rm EW}, \qquad\qquad  m_{q_B} = \frac{Y_{q_B}}{\sqrt{2}} v_B~,
\end{equation}
and we again expand $\tilde\sigma$ and $\tilde h$ using
(\ref{rotation}), we will obtain the fermion couplings to the physical
scalar eigenstates. As before, the couplings of the $\sigma$ to the
visible $A$ sector are suppressed by $\sin(\theta-\alpha)$, whereas to
the $B$ sector they are suppressed by $\cos(\theta-\alpha)$.
And viceversa, the couplings of the physical Higgs boson to the $A$
sector are suppressed by $\cos(\theta-\alpha)$, and to the $B$ sector
by $\sin(\theta-\alpha)$. In general, we see that we the couplings of
the scalar to the SM particles in the $A$ sector go like
\begin{eqnarray}
  g_\sigma^{\rm SM} = \sin(\theta-\alpha)\,g^{\rm SM},
  \label{sigmacoupling}
  \end{eqnarray}
  where  $g^{\rm SM}$ represents  the SM coupling of that particle to
  the Higgs boson in the SM.
  On the other hand, for the Higgs we have
  \begin{equation}
g_{h}^{\rm SM} = \cos(\theta-\alpha)\,g^{\rm SM}
\end{equation}
We all this information we are ready to study the production and
decays of the $\sigma$ at the LHC.

\section{Phenomenology of the Radial State at the LHC}
\label{sec:pheno}
In this section we consider the phenomenology of the radial state
$\sigma$ in composite Higgs models as well as in the Twin Higgs. 
We will first obtain the current bounds from the LHC at the end of Run
2 by using the data for searches for heavy scalar particles decaying either
to $ZZ\to 4\ell$, or to $hh$, a pair of Higgs bosons. We will also
consider the reach of the HL-LHC in these channels in order to asses
the possibility of discovery of the radial state in associated with a
pNGB Higgs scenario. The aim is to see if, in addition to the
precision measurements of the Higgs boson couplings, the direct searches  
for these scalar states can be an important complement in exploring
the nature of the Higgs sector in these types of theories.

\begin{figure}[h]
   \center 
  %\hspace{0.4cm}
    \includegraphics[scale=0.35]{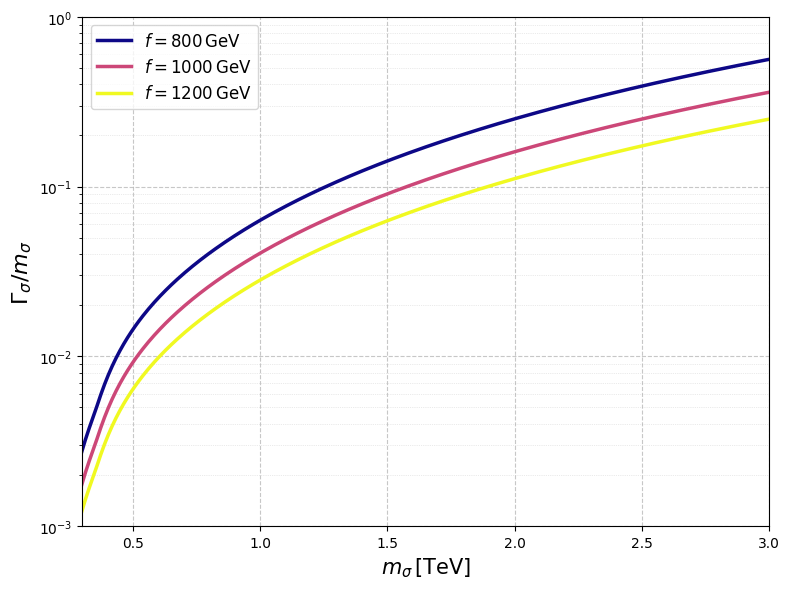} \includegraphics[scale=0.28]{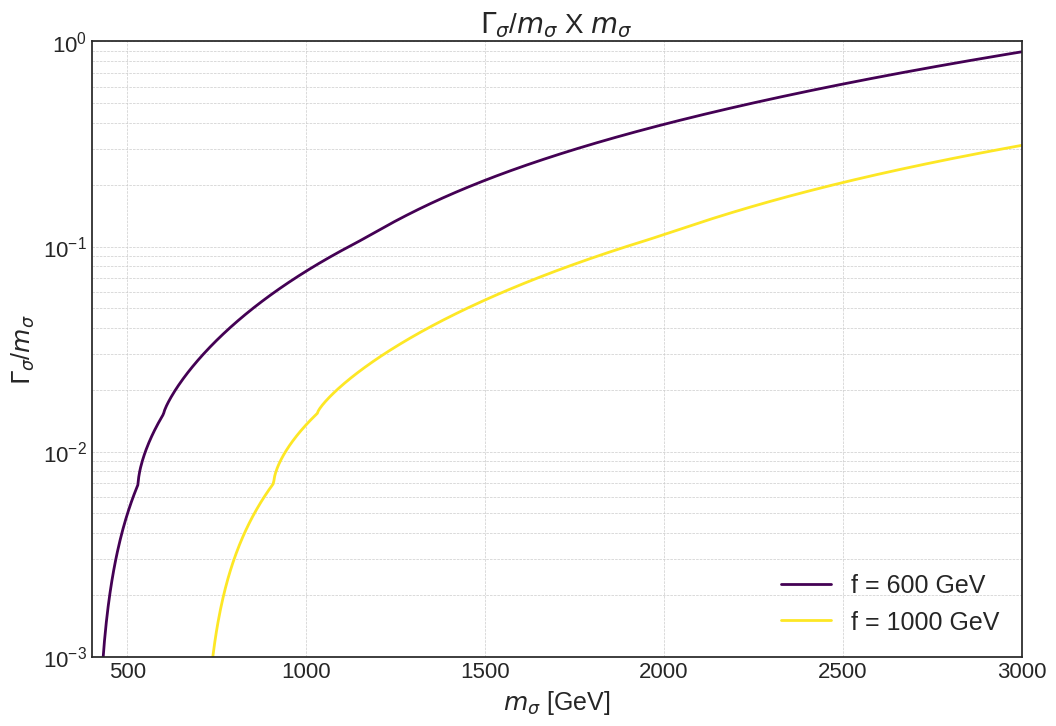} 
    \caption{The $\sigma$ total width to mass ratio vs. $m_\sigma$ for the MCHM (left
      panel) and the THM (right panel).  
    }
    \label{fig:widths}
\end{figure}

In what follows, we will compute the s-channel production of the
radial state in gluon fusion, followed by its decay into the
final states $X=ZZ, hh$, which dominate the searches. In our
calculation we will make use
of the narrow width approximation (NWA), which corresponds to
\begin{equation}
\sigma(gg\to \sigma\to X) = \sigma(gg\to\sigma)\,{\rm Br}(\sigma\to
X)~,
\label{nwa}
\end{equation}  
where the branching ration is
\begin{equation}
  {\rm Br}(\sigma\to X) = \frac{\Gamma(\sigma\to X)}{\Gamma_\sigma}~,
  \label{brdef}
\end{equation}  
and $\Gamma_\sigma$ is the $\sigma$'s total width.   
In the context of the s-channel production of a scalar such as the
radial excitation $sigma$, corrections to the NWA can be of importance
for large values of $\Gamma_\sigma$. The sources  of interference of
interference effects
~\cite{Fuchs:2014ola,Feuerstake:2024uxs} would be dominated by the SM
double Higgs production, for $X=hh$, or direct $ZZ$
production. However, these processe are mostly occurring at
considerably smaller values of $\sqrt{\hat s}$, which typically
suppresses the effects by $m_h^2/m_\sigma^2$ or $M_Z^2/m_\sigma^2$, as
long as $\Gamma_\sigma$ is not too large.
On the other hand, in these models $\Gamma_\sigma$ can be quite
large. This is  potentially an important correction to
(\ref{nwa})~\cite{Berdine:2007uv}. To consider the first corrections
to the NWA coming from a finite $\Gamma_\sigma$, we write
\begin{equation}
\sigma(gg\to\sigma\to X) = \sigma(gg\to \sigma)\,\int_{p^2_{\rm
    min.}}^{p^2_{\rm max.}} \frac{dp^2}{2\pi}
\frac{2\sqrt{p^2}}{|p^2-m^2_\sigma+i\Gamma_\sigma m_\sigma|^2}
\,\Gamma(\sigma\to X)~,
\label{nwa-corr}
\end{equation}  
where we assume that in the integrand $\Gamma_\sigma$ is constant, a
good approximation if the region  of integration is small. We assume
that the integration above takes place in the interval $[m_\sigma
-\Gamma_\sigma)^2, (m_\sigma+\Gamma_\sigma)^2]$. Performing the
integral and expanding in $\Gamma_\sigma/m_\sigma$, we obtain
\begin{equation}
\sigma(gg\to\sigma\to X) = \sigma(gg\to \sigma)\,{\rm Br}(\sigma\to
X)\,\left(1-\frac{\Gamma_\sigma}{\pi m_\sigma} +\cdots\right)~.
\label{nwa_1}
\end{equation}  
Equation (\ref{nwa_1}) gives the leading correction to the NWA for
finite width $\Gamma_\sigma$. We see that, as it may be intuitively
expected, a finite width leads to a reduction with respect to the NWA
expectation.Thus, for instance in order to have a reduction  of $10\%$
or less in the cross section, we must have
\begin{equation}
\frac{\Gamma_\sigma}{m_\sigma}\lesssim 0.30. 
\end{equation}
In Figure~\ref{fig:widths}, we show the total width  $\Gamma_\sigma$
for  the various
scenarios studied. The width of the radial state in the
MCHMs studied here tends to be larger due to the fact that the
dominant contributions go like $g_*^3$, as it can be seen in (\ref{sigmawidths}),
whereas for the THM the dominant modes   are further suppressed 
 by $\sin(\theta-\alpha)^2$,  as it can be inferred from (\ref{sigmacoupling}).  

 In the next section we will present constraints from data in the
NWA. However,  we should have in mind that for the corresponding point in
the cross section obtained using the NWA, one can obtain and estimate
of 
the reduction due to the width as given by (\ref{nwa_1})  in each
case.

\subsection{Bounds on the Composite Higgs Model Radial State}
\label{subsec:boundschm}
We first study the LHC bounds of the CHMs presented in
Section~\ref{sec:chmradial}. The production cross section of the
$\sigma$ in these models is dominated by gluon fusion. Thus, we will
consider this production channel exclusively. The loop diagrams
resulting in  the effective $\sigma-g-g$ coupling defined in  (\ref{leffggsigma}) 
receive contributions from the top quark as well as from the fermion
resonances characteristic of the specific model. These contributions
are given  in (\ref{toploop}) and (\ref{resloop}). To obtain the
prediction for $pp\to \sigma\to X$, with $X=ZZ$ or $hh$, we make use
of the $N^3LO$ gluon fusion predictions as given in
\cite{Anastasiou:2016hlm,LHCHiggsCrossSectionWorkingGroup:2016ypw} and
updated in \cite{cernyellowreport4update}. The production cross
section is given by
\begin{equation}
  \sigma^{gg}(pp\to \sigma) = \sigma^{gg}(pp\to h_{\rm
    SM})\,\left|\frac{c^t_{gg} +
      c^\psi_{gg}}{c^t_{gg}}\right|^2\,\left(\kappa^t_\sigma\right)^2~,
  \label{sigmaprodxsec}
\end{equation}
where $\sigma^{gg}(pp\to h_{\rm SM})$ is production cross
section at $\sqrt{s} = 13~$TeV,  in gluon fusion for  a heavy SM Higgs boson as computed in the references above to $N^3LO$
accuracy and using PDF4LHC\_15\_nnlo\_30 parton distribution functions,
and
\begin{equation}
  \kappa^t_{\sigma} = \frac{v}{f}~,
  \label{topsigma}
\end{equation}
reflects the fact that  the coupling of $\sigma$ to the top quark
differs from that of the SM Higgs. We use the cross section
above to compare with the searches performed at the LHC  in Run 2
at $\sqrt{s} = 13~$TeV. In all cases we use the narrow width
approximation, which is also assumed in the experimental cross section
bounds presented below. Thus, with the partial widths given in
(\ref{sigmawidths}) and the production cross section given in
(\ref{sigmaprodxsec}) above we are ready to extract the current bounds
from the LHC.
\begin{figure}[h]
   \center 
  %\hspace{0.4cm}
    \includegraphics[scale=0.70]{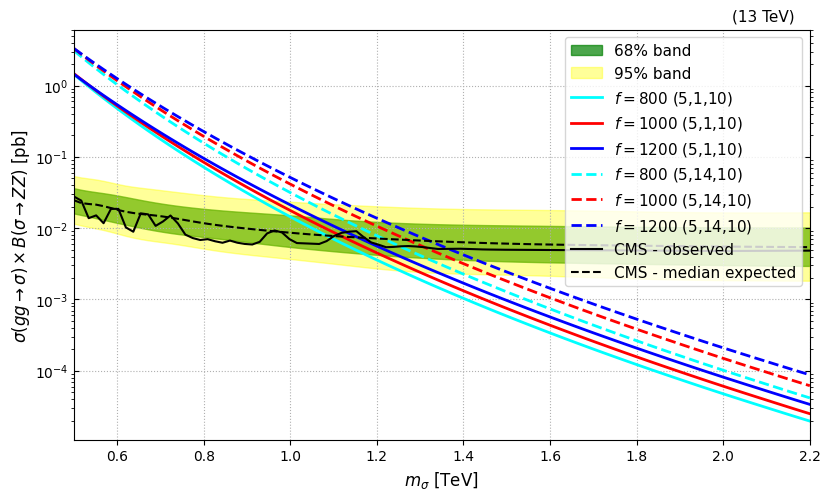}
    \caption{Production cross section of $gg \to \sigma\to ZZ$. The
      solid (dashed) line corresponds to the observed (expected)
      limit. The
      bands refer to the $1\sigma$ and $2\sigma$ limits.  From data by CMS
      in Run 2 with $138~fb^{-1}$ integrated
      luminosity~\cite{CMS:2024vps}.  The colored solid and dashed
      lines correspond to various predictions in the MCHM, for various
      values of the symmetry breaking scale $f$. See text for details. 
    }
    \label{fig:lhc_1}
\end{figure}
We consider two channels: the so called golden channel $\sigma\to
ZZ\to 4\ell$, and the double Higgs channel $\sigma\to hh$, where the
Higgs bosons can decay in a variety of ways.
In Figure~\ref{fig:lhc_1} we show the CMS
bounds on the production gross section $pp(gg) \to\sigma\to ZZ\to
4\ell$ from gluon fusion obtained with a luminosity of $138 fb^{-1}$
during Run 2~\cite{CMS:2024vps}.
The
bounds are compared with the predictions for the cross sections for
the radial mode, shown by the colored solid lines for the fermion
representation in $\mathbf{MCHM}_{\bf 5,1,10}$, and by the colored
dashed lines  for the case of $\mathbf{MCHM}_{\bf 5,14,10}$, both for
various values of the symmetry breaking scale $f$. We conservatively
consider $f\geq 800~$GeV, even though the actual bounds on this scale
coming from the measurements of the Higgs boson coupling are somewhat
smaller~\cite{Sanz:2017tco}. The predictions have a mild dependence of
modest variations of $f$.
On the other hand, the choice of fermion
representation has a larger impact on the results. 
We can extract bounds on $m_\sigma$ from
the displayed dataset. We see that the $2\sigma$ bounds are 
\begin{eqnarray}
  \mathbf{MCHM}_{\bf 5,1,10}:  \qquad\qquad m_\sigma \geq (930-980)
  {\rm GeV} \nonumber\\
 \mathbf{MCHM}_{\bf 5,14,10}:  \qquad\qquad m_\sigma \geq (1.03-1.13)
  {\rm TeV} 
  \label{lhc_bounds_zz}
\end{eqnarray}
depending on the values of $f$: $800~ {\rm GeV}, 1000~{\rm GeV},
1200~{\rm GeV}$.  
The bounds on the radial mode mass are more stringent in the case of
$\mathbf{MCHM}_{\bf 5,14,10}$, since the higher fermion representation
chosen here enhances the $\sigma$ production via extra contributions
to the loop diagram responsible for  $gg\to\sigma$.
\begin{figure}[h]
   \center 
  %\hspace{0.4cm}
    \includegraphics[scale=0.70]{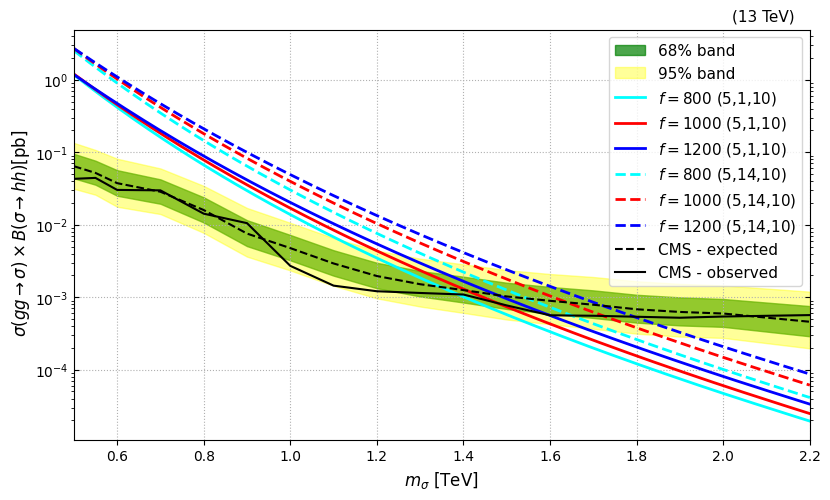}
    \caption{Combination of bounds on the cross section of $gg \to
      \sigma\to hh$ from ~\cite{CMS:2024vps}, with  $138~fb^{-1}$ integrated
      luminosity.   The
      solid (dashed) line corresponds to the observed (expected)
      limit. The
      bands refer to the $1\sigma$ and $2\sigma$ limits.  The colored solid and dashed
      lines correspond to various predictions in the MCHM, for various
      values of the symmetry breaking scale $f$. See text for details. 
    }
    \label{fig:lhc_2}
\end{figure}
On the other hand, when we  consider the $\sigma\to hh$ decay channel
the constraints on $m_\sigma$ are tighter. The CMS collaboration has published the results for
$138 fb^{-1}$ in this channel~\cite{CMS:2024phk}. As we see
in Figure~\ref{fig:lhc_2}, this channel has become quite competitive with the
$ZZ\to4\ell$. For lower masses, the bounds in Figure ~\ref{fig:lhc_2}
come from the $hh\to b\bar b \gamma\gamma$ channel, whereas for
masses above $1~$TeV the great progress made in the channel $hh\to
b\bar b b\bar b$, with pairs of b's merged, dominates. 

The $hh$ final state combination has surpassed the $ZZ\to 4\ell$
bounds. We see that we obtain
\begin{eqnarray}
  \mathbf{MCHM}_{\bf 5,1,10}:  \qquad\qquad m_\sigma \geq (1.0-1.25)
  {\rm TeV} \nonumber\\
 \mathbf{MCHM}_{\bf 5,14,10}:  \qquad\qquad m_\sigma \geq (1.3-1.5)
  {\rm TeV} 
  \label{lhc_bounds_hh}
\end{eqnarray}
depending on the values of $f$: $800~ {\rm GeV}, 1000~{\rm GeV},
1200~{\rm GeV}$.  
We see that the $hh$ channel is quite constraining on $m_\sigma$
already wth the Run 2 dataset, more so than the golden channel. This
will carry over to the HL-LHC projections, as we will see below.

\subsection{Bounds on the Twin Higgs Radial State}
\label{subsec:boundsthm}
\begin{figure}[h]
   \center 
  %\hspace{0.4cm}
    \includegraphics[scale=0.40]{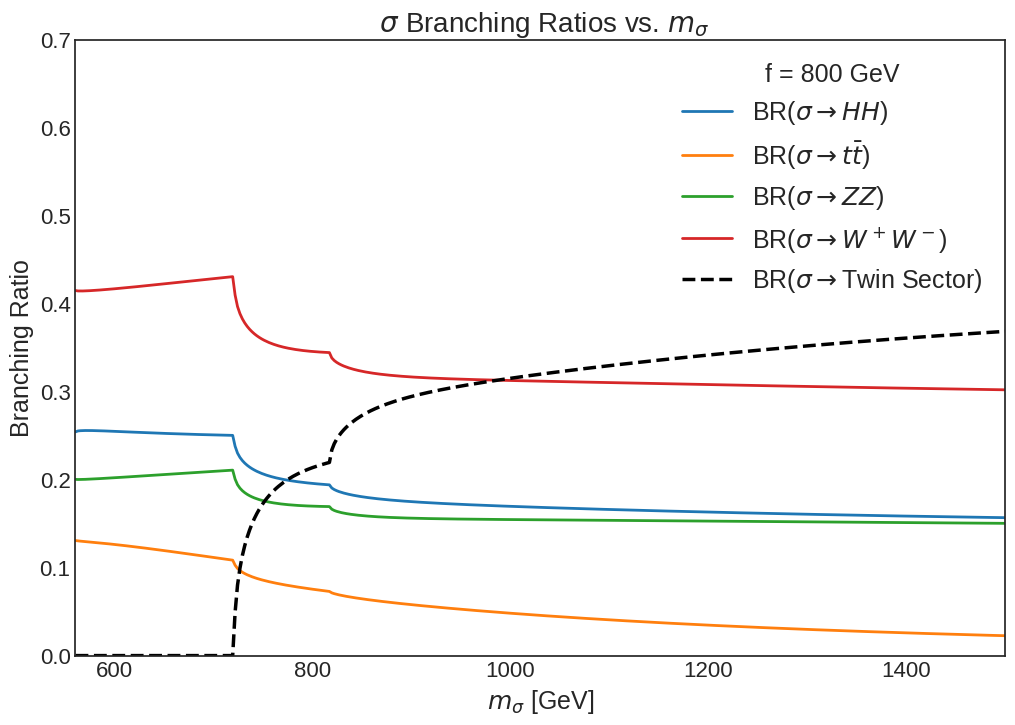}
    \caption{Branching fractions of $\sigma$ decays for $f=800~$GeV. The dashed line
      corresponds to the invisible $\sigma$ width. 
    }
    \label{fig:brsigmatwin}
  \end{figure}

The radial mode in the MTH is produced through gluon fusion, just as
the case in CHMs. On the other hand, the loop only contains the SM top
quark. Once again, assuming the narrow width approximation, we compute
the cross section for $pp\to \sigma\to X$, with $X=ZZ, hh$ as
\begin{equation}
  \sigma(pp\to \sigma \to X) = \sigma(pp\to \sigma ) * {\rm Br}(\sigma\to
  X),
  \label{crossbr}
  \end{equation}
where we make use of the $N^3LO$ production cross section as computed
in \cite{Anastasiou:2016hlm,LHCHiggsCrossSectionWorkingGroup:2016ypw,cernyellowreport4update}, appropriately rescaled by a factor of
$\sin^2(\theta-\alpha)$ from (\ref{sigmacoupling}). The branching fractions for
$\sigma$ as illustrated in Figure~\ref{fig:brsigmatwin}.
 Once $m_\sigma$ is large enough to decay to twin gauge bosons, which
 are heavier than the SM ones roughly by a factor of $f/v_{\rm EW}$,
 the invisible width dominates. Among the visible channels, the
 $W^+W^-$ is largest, followed by $hh$ and $ZZ$, which are used for
 the searches.

 The onset of the visible width, plus the fact that only the top quark
 contributes in the loop~\footnote{The top partners in the TH models
   do no carry color so they do not contribute to $pp\to\sigma$.} make
 the cross sections for the production of the radial state in the twin
 Higgs model considerably smaller than for the case of MCHMs studied
 earlier. In particular, when considering the Run 2 dataset with $138
 fb^{-1}$ of integrated luminosity,the constraints are considerably
 weaker on $m_\sigma$.  
A bound on $f$  can be derived from the measurements of the Higgs
boson couplings ~\cite{CMS:2022dwd}. The coupling measurements
coincide with the SM expectations to $1\sigma$ accuracy, allowing
deviations between $10\%$ $20\%$.
This translates into the bounds
\begin{equation}
f > 675~{\rm GeV} (610~{\rm GeV})~,
\end{equation}
where the first number corresponds to allowing  $10\%$ deviations, and
the number in parenthesis allows for a $20\%$ deviation.  
Using
(\ref{msigmabound}) we have that the radial state mass in the TH model
should satisfy
 \begin{equation}
   M_{\sigma}\gtrsim {\rm 475~GeV} (425~{\rm GeV}).
   \label{msigmarun2}
 \end{equation}
The Run 2 dataset of $138 {\rm fb}^{-1}$ luminosity is not
constraining $m_\sigma$ for the values of $f$ allowed by the Higgs
coupling measurements. Thus, we should consider (\ref{msigmarun2}) the
lower bound on the mass of the radial state as of the end of Run. We
will see below that the HL-LHC will have a better reach on
$m_\sigma$.

\section{Reach at the HL-LHC}
\label{sec:future}
In this section we examine the reach of the HL-LHC for the radial
state in both CHMs as well as the TH.
The ATLAS and CMS collaborations project their sensitivity for the
production of a scalar in the s channel, decaying to a variety of
modes, at $\sqrt{s}=14~$ TeV and with a benchmark integrated luminosity of 
$3000 ~{\rm fb}^{-1}$~\cite{ATLAS:2025eii}. We focus once again on the
$\sigma\to ZZ$ and $\sigma\to hh$ decay modes.

\begin{figure}[h]
   \center 
  %\hspace{0.4cm}
    \includegraphics[scale=0.35]{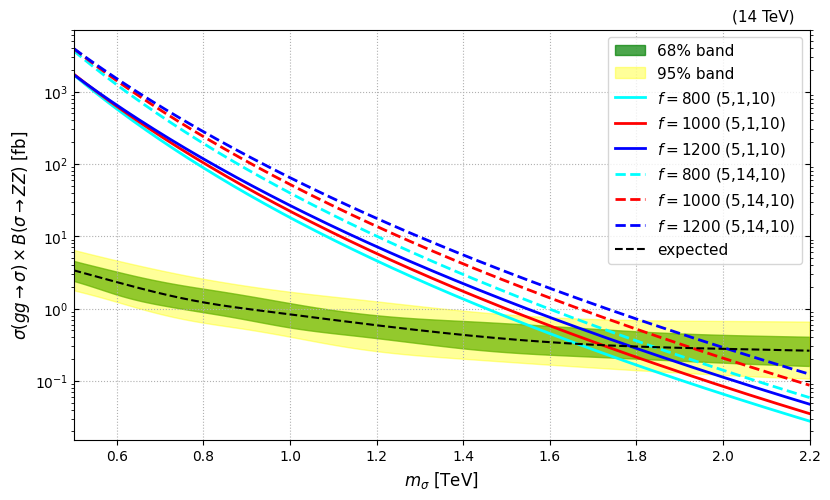} \includegraphics[scale=0.35]{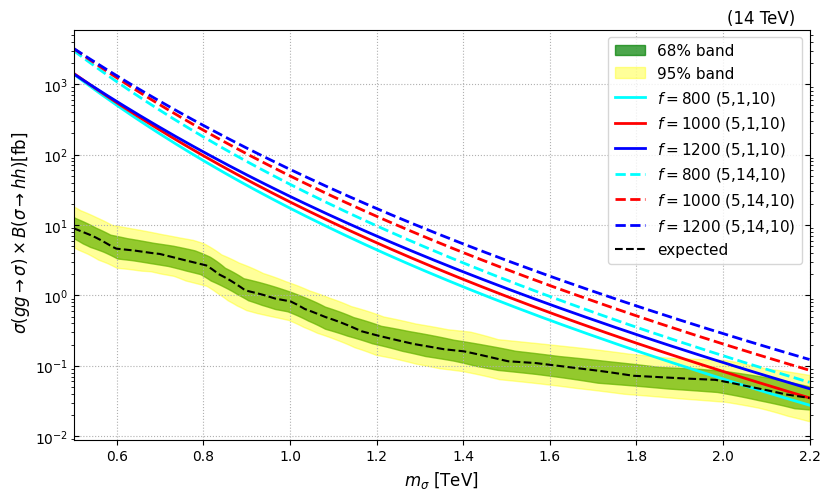} 
    \caption{Combination of ATLAS and CMS projections for 
     $pp\to \sigma\to X$ from ~\cite{ATLAS:2025eii}, with  ${\cal L} =
     3000~{\rm fb}^{-1}$ integrated
      luminosity. Left panel: $X=ZZ$ combination for
      $Z\to\ell^+\ell^-$. Right panel: $X=hh$, with $hh\to b\bar b,
      \gamma\gamma, 4b's ({\rm merged})$. 
      The solid (dashed) line corresponds to the observed (expected)
      limit. The
      bands refer to the $1\sigma$ and $2\sigma$ limits.  The colored solid and dashed
      lines correspond to various predictions in the MCHM, for various
      values of the symmetry breaking scale $f$. See text for details. 
    }
    \label{fig:hl-lhc_1}
\end{figure}

For the MCHMs considered here, the potential reach of the HL-LHC  can
be inferred from Figure~\ref{fig:hl-lhc_1}.
We see that the more constraining bounds appear to be on the
$\sigma\to hh$ channels. Likely this reflects the great progress being
made in these channels in anticipation of the measurement of the Higgs
boson self coupling through $h^*\to hh$ at the HL-LHC. In the right
panel of the Figure, the region for $m_\sigma<1~$TeV is still dominated
by $hh\to b\bar b \gamma\gamma$. But for $m_\sigma\geq 1~$TeV, the
bounds are dominated by $hh\to b\bar b b\bar b$, where pairs of b jets
are merged. This results in the potential reach of the HL-LHC on
$m_\sigma$ coming from the $\sigma\to hh$ channels.   From
Figure~\ref{fig:hl-lhc_1} we obtain
\begin{eqnarray}
  \mathbf{MCHM}_{\bf 5,1,10}:  \qquad\qquad m_\sigma \geq (1.82-1.98)
  {\rm TeV} \nonumber\\
 \mathbf{MCHM}_{\bf 5,14,10}:  \qquad\qquad m_\sigma \geq (2.04-2.24)
  {\rm TeV} 
  \label{hl-lhc_bounds_hh}
\end{eqnarray}
with the intervals corresponding to the various values of the scale
$f$ as specified in the Figure.   
This is to be compared with the current bounds from Run 2 in
(\ref{lhc_bounds_hh}). We can also compare these reach on $m_\sigma$
above with the one obtained from the $\sigma\to ZZ$ in the left panel
of Figure~\ref{fig:hl-lhc_1}, where the bounds range from $1.45~$TeV
to $1.78~$TeV for the same variations of $f$ and the fermion
representations used in the right panel for $\sigma\to hh$. We 
conclude that the $\sigma\to hh$ wil have the best reach for
discovering or constraining the radial state of composite Higgs
models.
\begin{figure}[h]
   \center 
  %\hspace{0.4cm}
    \includegraphics[scale=0.24]{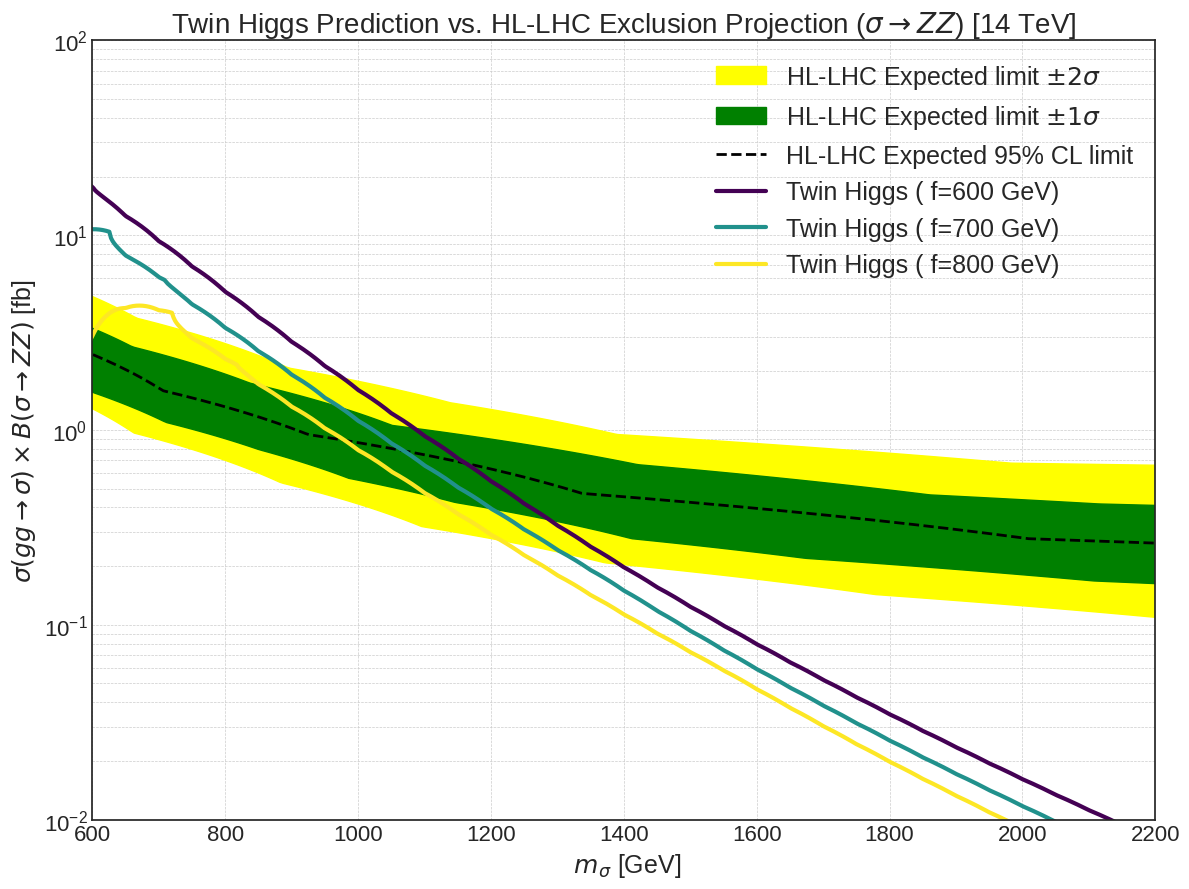} \includegraphics[scale=0.24]{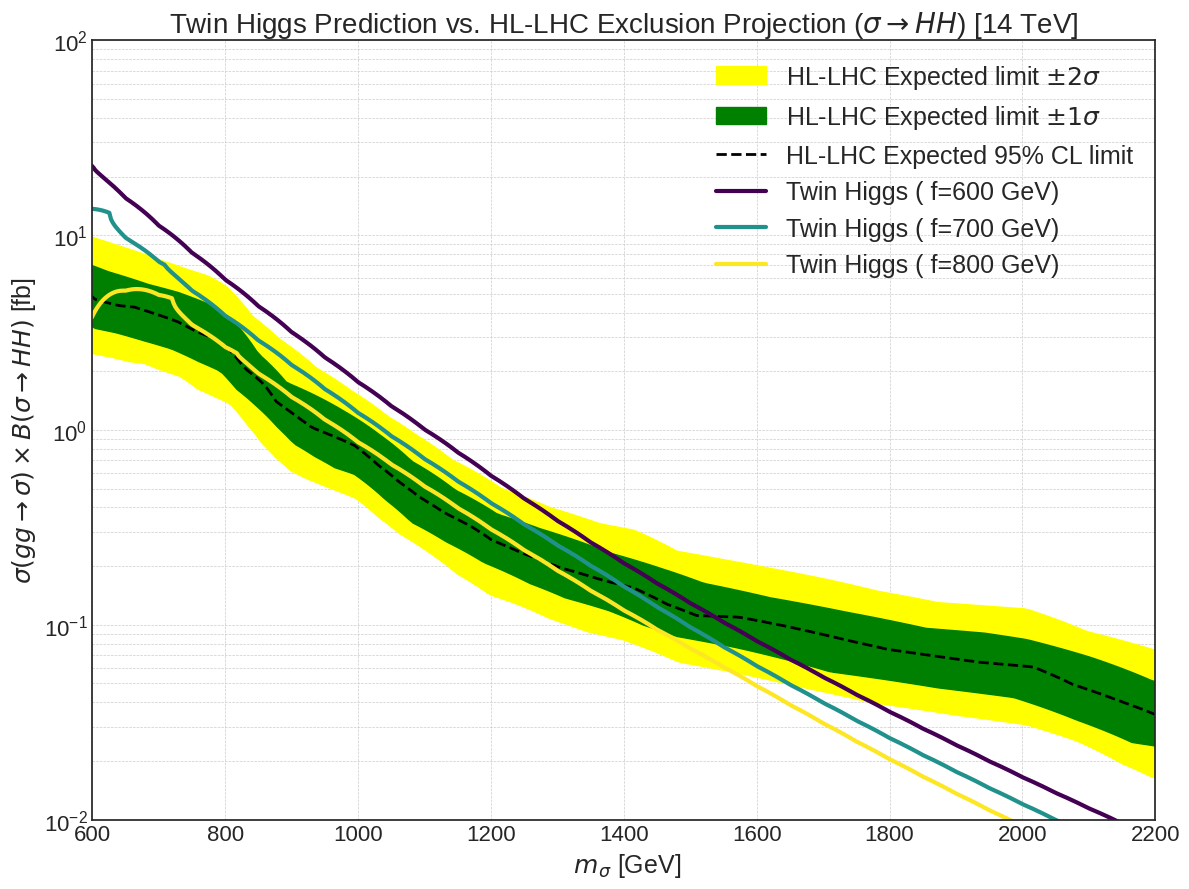} 
    \caption{Combination of ATLAS and CMS projections for 
     $pp\to \sigma\to X$ from ~\cite{ATLAS:2025eii}, with  ${\cal L} =
     3000~{\rm fb}^{-1}$ integrated
      luminosity. Left panel: $X=ZZ$ combination for
      $Z\to\ell^+\ell^-$. Right panel: $X=hh$, with $hh\to b\bar b,
      \gamma\gamma, 4b's ({\rm merged})$. 
      The solid (dashed) line corresponds to the observed (expected)
      limit. The
      bands refer to the $1\sigma$ and $2\sigma$ limits.  The colored
      solid lines correspond to different values of the symmetry
      breaking parameter $f$. See text for details. 
    }
    \label{fig:hl-lhc_2}
  \end{figure}
  
In the case of the THM radial state, the HL-LHC will have reach for
direct searches for the first time. In Figure~\ref{fig:hl-lhc_2} we
show the predictions for the $ZZ\to 4\ell$ (left panel) and for the
combination of all  $hh$ decay modes. The projections for
the  data with ${\cal L}=3000 ~{\rm fb}^{-1}$ are from
Ref.~\cite{ATLAS:2025eii}.
Although it appears the $hh$ mode
combination outperforms the $ZZ$ ``golden channel'', this is only for
the case of the symmetry breaking scale $f=600~$GeV, which is quite at
the edge of what is allowed by the current measurements of Higgs
couplings, as discussed in the previous section. 
On the other hand, when considering $f=700$~GeV, consisten with
deviations of Higgs couplings of up to $10\%$, the bes bound comes
from the $ZZ\to 4\ell$ channel. So the HL-LHC reach is obtain from the 
 $hh$ ATLAS-CMS combination predictions for  $f=600$~GeV: 
\begin{equation}
  m_\sigma \geq 1.2~ {\rm TeV}~.
 \label{twinhllhcbound_1} 
\end{equation}
On the other hand, making use of the more conservative values
$f=700$~GeV, the best reach appears to be in the $ZZ\to 4\ell$ channel
and is
\begin{equation}
  m_\sigma \geq 850~ {\rm GeV}~.
 \label{twinhllhcbound_2} 
\end{equation}
The bound projections in  (\ref{twinhllhcbound_1}) and  (\ref{twinhllhcbound_2}) indicate that the HL-LHC will
have reach to discover the THM radial state, perhaps the only accessible new
particle in this set of models solving the hierarchy problem. 
This values for the radial state mass are still consistent with a fairly
narrow state, indicating that there will still be room for discovering
this particle were the HL-LHC to either increase the total luminosity
accumulated beyond $3000~{\rm fb}^{-1}$, or the techniques used in the
relevant final states were to be bested. Both of these possibilities
seem realistic. Particularly, the role of the $hh\to b\bar b b \bar b$
decay modes to make the $hh$ channel the most sensitive, as well as
the its importance in testing the SM quartic coupling, suggests that
once the HL-LHC is running the performance of this channel could show
significant progress.

Of course, the bounds on the parameter $f$ will be much stronger at
the end of the HL-LHC due to the projected increased precision in the
measurements  of the Higgs couplings. For instance, combining the
projections of ATLAS and CMS with $3000~{\rm fb}^{-1}$ integrated
luminosity~\cite{ATLAS:2025eii}, the gauge boson couplings will be known with a $1.6\%$
precision. If no deviations from the SM couplings are observed, this
would point to a bound on the THM symmetry breaking scale of
$f>975~$GeV. Then, as can be seen  in Figure~\ref{fig:hl-lhc_2}, this
scale would render the radial mode too heavy to be observed directly
at the HL-LHC. Then, if a new scalar is observed in the high
luminosity run it is important to correlate its mass, as defined by
$f$, with the corresponding deviations of the Higgs couplings to gauge
bosons and fermions.

\section{Conclusions} 
\label{sec:conc}
Extensions of the SM where the Higgs is a pNGB provide an insight into
the nature of the Higgs sector and a solution to the little hierarchy
problem. 
So far all the experimental constraints on this
scenario for the Higgs boson have concentrated on the bounds on its
couplings, as well as the searches for vector and vector-like fermion resonances
present in some of these models. In this paper we have shown that the LHC has
reach to search for the radial excitation that is present in
these scenarios, whether in  composi Higgs models  or the twin Higgs model.

In the case of composite Higgs models, we have obtained current bounds
on the mass of the radial state, $m_\sigma$, from Run 2 data. This is
shown in (\ref{lhc_bounds_hh}), the $2\sigma$ bound derived from the
$\sigma\to hh$ channels, using ${\cal L}=138~{\rm fb}^{-1}$ collected
from ATLAS and CMS in Run 2. The bounds are for the MCHM, where we
used specific fermion representations consistent with the existence of
a narrow radial mode that can be seen at the LHC. These bounds 
range from $m_\sigma\leq 1.0~$TeV to  $1.5~$TeV, depending on details
of the MCHM model. It shows that the LHC has still reach for new
scalar states, and that there is another way to get at composite Higgs
models that complements the higher precision measurements of Higgs
couplings that will surely dominate the future at the HL-LHC.

We have also shown that the $\sigma\to hh$  channels are very
competitive for performing this search, and that sometimes  they
perform better than 
the $\sigma\to Z\to 4\ell$ 
This becomes clearer as we study the reach of the
HL-LHC in $m_\sigma$ in the MCHM that we study. As the mass of the
radial state becomes larger than $1$~TeV, the $hh\to b\bar b b\bar b$
channel takes over the constraints. It is apparent than the
improvement in these channels plays a crucial role in driving the
reach of the HL-LHC for the radial state. This is clear from
Figure~\ref{fig:hl-lhc_1} (right panel), from which we obtained the
reach in $m_\sigma$ given in (\ref{hl-lhc_bounds_hh}), stretching
beyond $2.2~$TeV.

For the case of the THM, we have seen that the LHC Run 2 dataset is
unable to obtain a better bound that what it is given by considering
the Higgs couplings constraints, as shown in (\ref{msigmarun2}).
On the other hand, we have shown that the HL-LHC will have reach
beyond this, as shown in (\ref{twinhllhcbound_1}), which corresponds to
$f=600~$GeV and (\ref{twinhllhcbound_2}) for $f=700$~GeV, values of the symmetry breaking scale in the
THM that are compatible with the current Higgs coupling bounds being
$20\%$ and $10\%$ deviated from the SM predictions, respectively. .
On the other hand,
the reach of the HL-LHC in the THM is not as  good as for the radial
state in MHCMs. This is  the result of the very nature of the THM,
which is very difficult to observe experimentally. In fact, the radial
state in the THM is likely to be the only state to be directly
observable, and then its pursuit should be central in testing this
highly competitive solution to  the hierarchy problem.

In sum, we have shown that the LHC can still search for and put interesting bounds on
the masses of new scalar particles that are well motivated by
extensions of the SM. In theories where the Higgs boson is a pNGB of a
spontaneously broken global symmetry, such as in CHM and the THM, the
radial excitation is an important ingredient. We have seen that the
current LHC data already puts important constraints in the case of the
MCHMs, whereas the HL-LHC will have an important reach in $m_\sigma$
for both CHM as well as the THM.

\acknowledgments
G.B. thanks the organizers of the Higgs@Capri 2025, as well as the
hospitality of the Department of Physics at Columbia University. 
The authors also acknowledge the financial support of FAPESP under grants
2019/04837-9, 2025/19987-7, 2024/03583-1, 2024/16149-8 and
2025/00161-1.

\bibliography{sigmarefs}
\bibliographystyle{JHEP}

\end{document}